\documentclass[journal]{IEEEtran}
\ifCLASSINFOpdf
\else
\fi
\usepackage{graphicx}
\usepackage{soul}
\usepackage{float}
\usepackage{colortbl}
\usepackage{graphicx} 
\usepackage{multirow}
\usepackage{makecell}
\usepackage{stfloats}
\usepackage{subfigure}
\usepackage{amsfonts,amssymb}
\usepackage{booktabs}
\usepackage{colortbl}
\usepackage[namelimits]{amsmath} 
\usepackage{amssymb}             
\usepackage{amsfonts}            
\usepackage{mathrsfs}            
\usepackage{cite} 
\usepackage{booktabs}   
\usepackage{etoolbox}
\makeatletter
\patchcmd{\@makecaption}
  {\scshape}
  {}
  {}
  {}
\makeatother
\usepackage{threeparttable}
\usepackage{color}
\usepackage{url}

\begin{document}

%
\title{End-to-end Transformer for Compressed Video Quality Enhancement}
%
%
%

\author{Li~Yu,
        Wenshuai~Chang,
        Shiyu~Wu
        and~Moncef~Gabbouj
\thanks{Preprint.}
\thanks{Li Yu, Wenshuai Chang, Shiyu Wu are with School of Computer and Software, Nanjing University of Information Science and Technology, Nanjing, China, and also with Engineering Research Center of Digital Forensics, Ministry of Education, Nanjing University of Information Science and Technology, Nanjing, China.}
\thanks{Moncef Gabbouj is with Department of Computing Sciences, Tampere University, Tampere, Finland.}


}

\maketitle

\begin{abstract}
Convolutional neural networks have achieved excellent results in compressed video quality enhancement task in recent years. State-of-the-art methods explore the spatio-temporal information of adjacent frames mainly by deformable convolution. However, offset fields in deformable convolution are difficult to train, and its instability in training often leads to offset overflow, which reduce the efficiency of correlation modeling. In this work, we propose a transformer-based compressed video quality enhancement (TVQE) method, consisting of \emph{Swin-AutoEncoder based Spatio-Temporal feature Fusion} (SSTF) module and \emph{Channel-wise Attention based Quality Enhancement} (CAQE) module. The proposed SSTF module learns both local and global features with the help of \emph{Swin-AutoEncoder}, which improves the ability of correlation modeling. Meanwhile, the window mechanism-based Swin Transformer and the encoder-decoder structure greatly improve the execution efficiency. On the other hand, the proposed CAQE module calculates the channel attention, which aggregates the temporal information between channels in the feature map, and finally achieves the efficient fusion of inter-frame information. Extensive experimental results on the JCT-VT test sequences show that the proposed method achieves better performance in average for both subjective and objective quality. Meanwhile, our proposed method outperforms existing ones in terms of both inference speed and GPU consumption.

\end{abstract}


%
\IEEEpeerreviewmaketitle

\section{Introduction}
\label{sec:introduction}
%
%
%
%

The multimedia industry is growing rapidly and consumers are expecting videos of higher quality. On the one hand, video is becoming the main form of information carrier in increasing applications, including remote education, telemedicine, live broadcasting, digital TV, video conference. On the other hand, the demand for video resolution is constantly increasing, from $1080p$ to $2K$, $3K$, $4K$, as well as $8K$. As a result, the extremely large amount of video data needs to be compressed by video compression algorithms, such as H.264/AVC~\cite{wiegand2003overview} and H.265/HEVC~\cite{sullivan2012overview}, to fit the
available storage and network bandwidth.
As the compression ratio increases, the encoder significantly reduces the bit rate while introducing undesirable artifacts that severely degrade the quality of experience (QoE). The introduced artifacts also impair the performance of downstream video-oriented tasks (e.g., action recognition~\cite{sun2022human,zhu2020comprehensive}, object tracking~\cite{xu2019deep,luo2021multiple} and video understanding~\cite{bertasius2021space,wu2021towards,wu2019long}). Therefore, it is vital to enhance the quality of compressed video.

 \begin{figure*}[htbp]
    \centering\includegraphics[width=18cm]{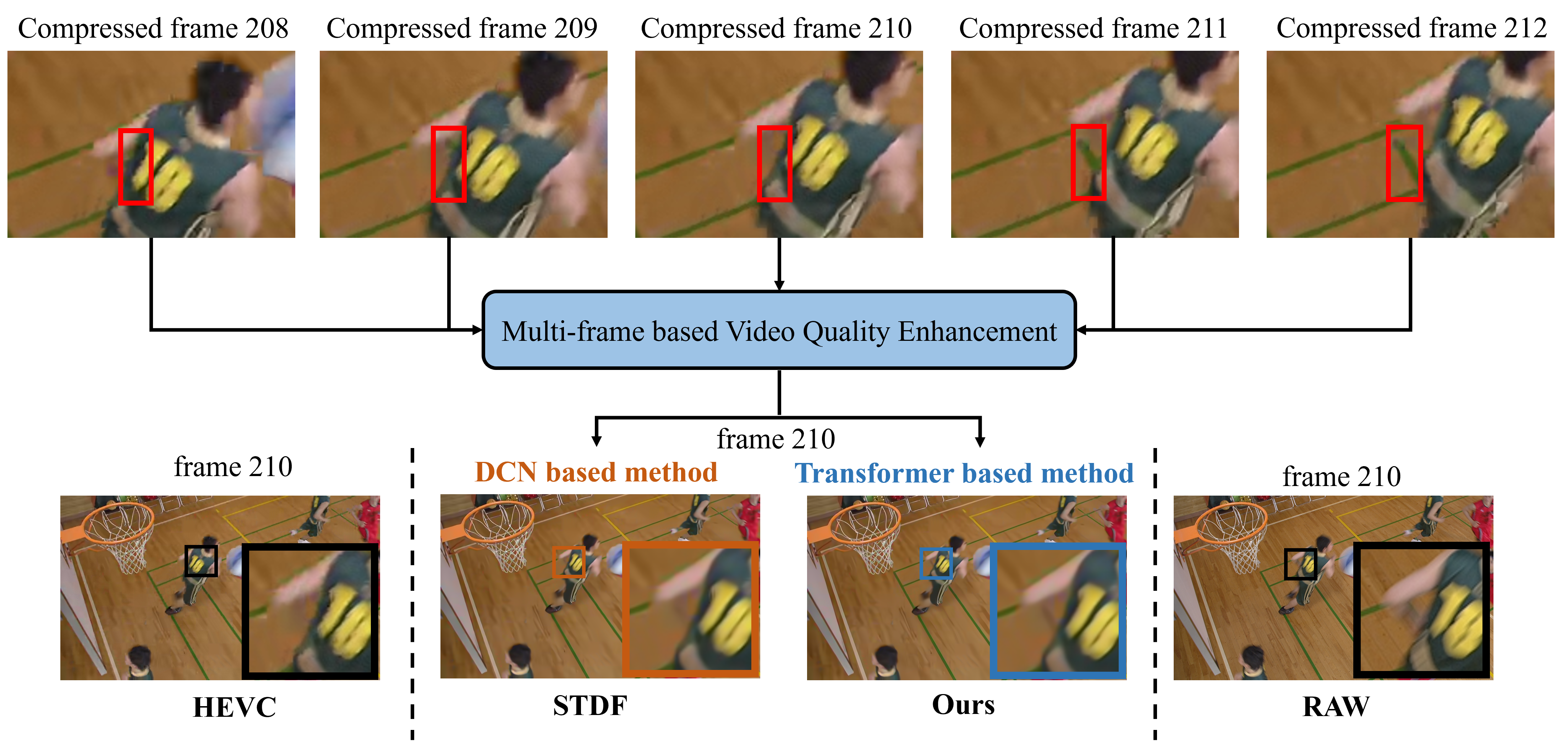}
\caption{An illustration showing the quality enhancement performance of our TVQE method, compared with DCN based method and HEVC (Class C, \emph{BasketballDrill}).}
\label{fig:example}
\end{figure*}

Convolutional neural networks (CNNs) have achieved milestones in the task of video quality enhancement (VQE). The CNN-based approaches can generally be classified into two categories: single-frame based methods~\cite{dong2015compression,guo2016building,li2017efficient,tai2017memnet,zhang2017beyond,chen2018dpw} and multi-frame based methods~\cite{yang2018multi,guan2019mfqe,xu2019non,deng2020spatio,zhao2021recursive,yang2021ntire}.
The single-frame based video enhancement method is equivalent to image enhancement, which explores the contextual information within the frame/image by CNNs to reduce compression artifacts and improve the visual quality. 
However, the temporal correlations between adjacent frames in the video are ignored, which severely limits the performance.
In multi-frame based methods, the temporal information between adjacent frames are explored. Since there are motions between adjacent frames, the inter-frame information cannot be used directly.
Some works use the optical flow to compensate the motion between frames. For example, \cite{yang2018multi,guan2019mfqe} used dense optical flow for motion compensation. However, the optical flow calculated from compressed video tends to be inaccurate. Thus, some work~\cite{deng2020spatio} utilizes deformable convolution (DCN) to capture the dependencies between multiple adjacent frames and the DCN-based approaches have made great progress in this task. However, deformable convolutional alignment modules are difficult to train, and its instability in training often leads to offset overflow, which ultimately reduces the efficiency of correlation modeling. Thus,~\cite{zhao2021recursive} proposed the Recursive Fusion (RF) module based on~\cite{deng2020spatio}, which saves the temporal information of previously enhanced video frames for correlation modeling, implicitly expanding the temporal range and achieving superior results. However, the RF module consumes huge GPU computing resources and slows down the inference speed.


In order to capture the long-range correlations between frames, we introduce vision transformer into the VQE task for its strong capability to learn long-range dependencies between image patch sequences and adaptability to image content. 
Since the computational complexity of the traditional vision transformer~\cite{dosovitskiy2020image} grows quadratically with the increase of image resolution, we use swin transformer~\cite{liu2021swin} in our work, along with the auto-encoder structure. The window-based swin transformer and multi-scale encoder-decoder structure with skip connections can improve the inference efficiency and reduce GPU consumption. Meanwhile, the swin transformer auto-encoder with skip connections facilitates the mining of spatio-temporal information, as well as correlation modeling of temporal information among multiple frames. 
As shown in Fig.~\ref{fig:example}, our method achieves better results than other methods. The compressed frame $210$ is enhanced with the information in frames from $207$ to $213$ (only frame $208$ to $212$ are drawn for illustration). It can be seen that the green line on the floor (below the athlete's arm) is totally or partially occluded in frames from $207$ to $209$, and becomes gradually clear from frame $210$ to $213$. In order to recover the green line in frame $210$, proper correlation among pixels should be modeled. The results show that our method achieves better recovery result in this region, which verifies its effective correlation modeling of temporal information among multiple frames. 
Besides the spatio-temporal information exploration, we employ \emph{Restormer}~\cite{zamir2022restormer} to calculate the channel attention, which enables efficient fusion of inter-frame information. 

The main contributions of this paper are summarized as follows:
\begin{itemize}
  \item We propose an compressed video enhancement transformer, which is the first work entirely based on a transformer-based architecture.
  \item 
  Compared to state-of-the-art DCN-based methods, our proposed method has better ability of long-range correlation modelling.
  \item Our proposed method calculates both spatio-temporal attention and channel attention, which effectively exploits the spatio-temporal information from multiple reference frames and achieves efficient fusion of the information.
  \item We conduct extensive experiments on the JCT-VT test sequence and demonstrate the effectiveness of the proposed method.
\end{itemize}

The rest of this paper is organized as follows. In Section~\ref{sec:related-work}, the vision transformer, deep learning based compressed video enhancement methods are reviewed. Section~\ref{sec:method} describes the task of VQE, the structure of the proposed TVQE method and the training scheme. Section~\ref{sec:experiment} presents the experiments and results. Finally, in section~\ref{sec:con}, we present the conclusions and future work.

 \begin{figure*}[htbp]
 \centerline{\includegraphics[width=18cm]{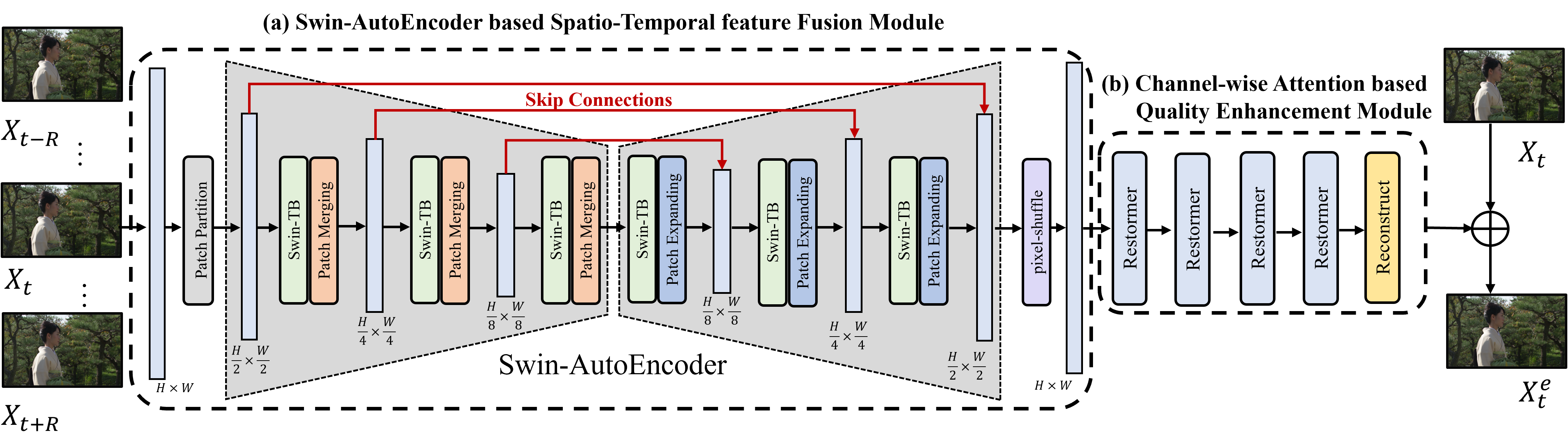}}
\caption{The framework of our proposed TVQE method, which consists of the \emph{Swin-AutoEncoder based Spatio-Temporal feature Fusion} (SSTF) Module and the \emph{Channel-wise Attention based Quality Enhancement} (CAQE) Module. The SSTF module is designed to exploit spatio-temporal correlation from multiple frames, where the \emph{Swin-AutoEncoder} (SAE), equipped with skip connections, is used. After SSTF, the information between channels in the feature map is further fused by the CAQE module, and finally generate the enhanced frame.
}
\label{fig:proposed}
\end{figure*}

\section{Related Work}
\label{sec:related-work}
In this section, we first review recent works on deep learning-based quality enhancement of compressed video, including single-frame based methods and multi-frame based methods. Then, a brief overview on vision transformer is provided.

\subsection{Single-frame based video enhancement method}
Single-frame video enhancement methods are considered as image enhancement. Earlier works~\cite{dong2015compression,li2017efficient,zhang2017beyond,chen2018dpw,galteri2017deep,guo2017one,liu2018non,yoo2018image,zhang2019residual} were mainly used for quality enhancement of JPEG compressed images. Specifically, AR-CNN~\cite{dong2015compression} first uses a convolutional neural network for image enhancement, and learns the nonlinear mapping between the original image and the compressed image with four convolutional layers. Subsequently, work such as~\cite{li2017efficient,tai2017memnet,zhang2017beyond,mao2016image,svoboda2016compression,zhang2017learning} proposed deeper networks to further improve the performance. With batch normalization and residual learning proposed, DnCNN~\cite{zhang2017beyond} effectively solves the problem of gradient disappearance in deep image enhancement networks. NLRN~\cite{liu2018non} and RNAN~\cite{zhang2019residual} proposed a residual non-local attention mechanism to capture long-range dependencies between pixels.
In addition to exploiting the information in the image spatial domain, methods such as~\cite{guo2016building,chen2018dpw,yoo2018image} exploited the relevant information in the frequency domain to further improve the subjective visual quality. In particular, \cite{dai2017convolutional,jin2018quality,wang2017novel,yang2018enhancing,yang2017decoder} also utilize the prior knowledge to improve the enhancement performance. 
For example, DS-CNN~\cite{yang2017decoder} and QE-CNN~\cite{yang2018enhancing} used different methods to deal with intra-frame coding (e.g., AI) and inter-frame coding (e.g., LDP, LDB, RA). In general, single-frame video enhancement methods ignore the temporal information in the video, thus the performance is severely limited.

\subsection{Multi-frame based video enhancement method}
Multi-frame video enhancement mainly utilizes the spatio-temporal information of multiple adjacent frames. Yang et al.~\cite{yang2018multi} firstly proposed Multi-Frame Quality Enhancement (MFQE $1.0$), which first uses SVM to divide high and low quality frames, and then use two adjacent high quality frames to perform motion compensation through optical flow and enhance the low quality frame. As an enhanced version of MFQE $1.0$, MFQE $2.0$~\cite{guan2019mfqe} proposed an end-to-end quality enhancement network, which pre-trained a bidirectional long short-term memory (BiLSTM) based model to detect peak quality frame (PQF).
The QE-subnet is also advanced by introducing the multi-scale strategy, batch normalization and dense connection. However, the video is compressed, and the compressed video can be severely distorted by various compression artifacts, so the estimated optical flow during motion compensation is often inaccurate and unreliable, resulting in ineffective quality enhancement. To this end, Deng et al. proposed a sliding window based method STDF~\cite{deng2020spatio}, which utilized deformable convolution to avoid explicit calculation of optical flow. This method innovatively proposed to perform feature alignment of moving objects on input multi-frame images through deformable convolution. Based on STDF, RFDA~\cite{zhao2021recursive} proposed the recursive fusion (RF) module, which not only utilized the reference frames within the current time window, but also exploits the temporal information of previously enhanced video frames. By implicitly expanding the time window, RFDA leveraged a larger range of temporal information for better spatio-temporal compensation. However, the computational complexity of RF module is huge.

\subsection{Vision Transformers}
Transformer is a deep neural network based on self-attention mechanism and parallel processing. Transformer~\cite{vaswani2017attention} emerged in the field of NLP. Its proposal solves the problems of recurrent network models, such as Long short term memory (LSTM)~\cite{he2016deep} and Gate recurrent unit (GRU)~\cite{chung2014empirical}. It cannot be trained in parallel and requires a lot of storage resources to memorize the entire sequence information. The successful application of Transformer in the field of NLP has made relevant scholars begin to discuss and try its application in the field of computer vision~\cite{dosovitskiy2020image,carion2020end}.
Image Transformer\cite{parmar2018image} was the first to migrate the Transformer architecture to the field of computer vision. Subsequently, Dosovitskiy et al.~\cite{dosovitskiy2020image} proposed the Visual Transformer (ViT), and ViT completely replaced the Transformer structure with the convolutional structure to deal with the classification task, and achieved results beyond CNN on extremely large-scale datasets~\cite{kolesnikov2020big,mahajan2018exploring,tou2019fixing,xie2020self}. However, the self-attention mechanism calculates the global similarity, and its computational complexity grows quadratically with the expansion of spatial resolution. To improve operational efficiency, an efficient and effective vision transformer called Swin Transformer was proposed in~\cite{liu2021swin}.
Based on the shift window mechanism, Swin Transformer achieves state-of-the-art performance in image classification~\cite{dosovitskiy2020image,wang2021pyramid,liu2021swin}, object detection~\cite{carion2020end,zhu2020deformable}, image segmentation~\cite{xie2021segformer,cheng2022masked}, video understanding~\cite{zhou2018end,zeng2020learning}, image generation~\cite{jiang2021transgan} and point clouds processing~\cite{zhao2021point,guo2021pct}. Zamir et al. proposed Restomer~\cite{zamir2022restormer}, which computes self-attention across channels rather than spatial dimensions, and its complexity grows linearly with image resolution. Thus, Restomer achieves state-of-the-art performance in large image restoration task. In this work, we use the Swin Transformer block as the basic unit to build a \emph{Swin-AutoEncoder} architecture with skip connections for aggregating the temporal information of multiple adjacent video frames. Then, we calculate the channel attention using \emph{Restormer} and efficiently fuse the temporal information to obtain the final result.


\section{Methodology}
\label{sec:method}

Given a compressed video consisting of $T$ frames $V=[X_{1},X_{2}...,X_{t},...,X_{T}]$, where $X_{t} \in \mathbb{R}^{H \times W}$ represents the compressed frame at time $t$, $H$ and $W$ are the height and width of $X_{t}$, the task of compressed video enhancement is to generate an enhanced video $V^{e} = [X^{e}_{1},X^{e}_{2}...,X^{e}_{t},...,X^{e}_{T}]$ from the input compressed video $V$.

The overall framework of the proposed method is shown in Fig.~\ref{fig:proposed}, which consists of two modules: (a) \emph{Swin-AutoEncoder based Spatio-Temporal feature Fusion} (SSTF) module and (b) \emph{Channel-wise Attention based Quality Enhancement} (CAQE) module. The SSTF module explores the spatio-temporal information from multiple frames by modeling the association of these frames with \emph{Swin-AutoEncoder} (SAE). After SSTF, the information between channels in the feature map is further fused by CAQE, and finally generate the residual of the enhanced frame.
For each compressed frame $X_{t}$, its $R$ preceding frames and $R$ succeeding frames are used to exploit correlated temporal information.
With the input $V_{t} = \left\{X_{t-R}, \ldots, X_{t}, \ldots, X_{t+R}\right\}$, the whole process can be expressed as:
\begin{eqnarray}
X_{t}^{m}&=&SA\left(V_{t}; \phi\right),\\
X_{t}^{e}&=&CA\left(X_{t}^{m} ; \varphi\right) + X_{t},
\label{equ:f1}
\end{eqnarray}
where $X_{t}^{e}$ is the output, $SA$ denotes the process of SSTF, and $CA$ denotes the process of CAQE. $\phi$ and $\varphi$ represent the parameters to be learned in the SSTF and CAQE modules, respectively. Finally, residual learning is used to improve the training efficiency.

 \begin{figure*}[htbp]
 \centerline{\includegraphics[width=18cm]{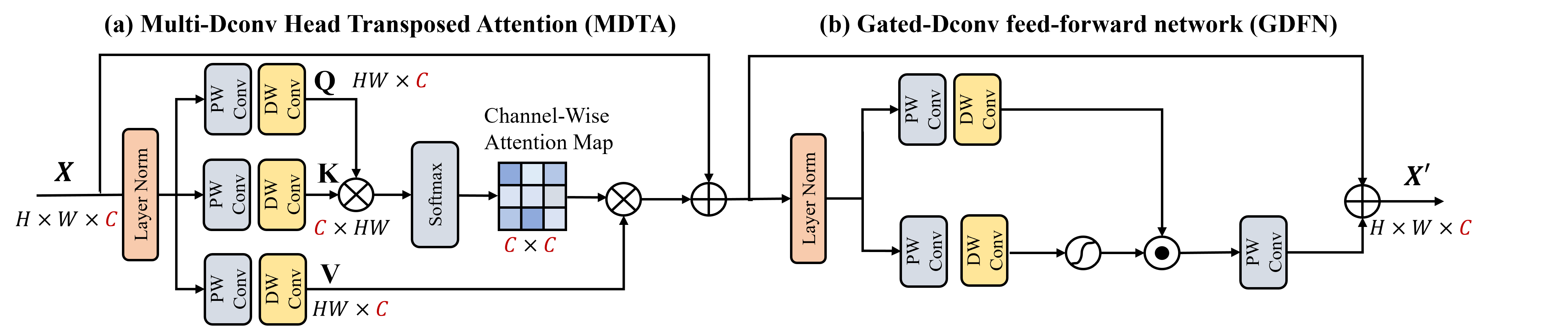}}
\caption{Architecture of \emph{Restormer}. It consists of (a) \emph{Multi-Dconv Head Transposed Attention} (MDTA) and (b) \emph{Gated-Dconv Feed-Forward Network} (GDFN). MDTA calculates channel-level attention and GDFN performs feature transformation by GELU to enrich feature representation.}
\label{fig:atten}
\end{figure*}

\subsection{Swin-AutoEncoder based Spatio-Temporal feature Fusion (SSTF)}
The SSTF module consists of the \emph{Patch Partition layer}, \emph{Swin-AutoEncoder} and \emph{Pixel Shuffle layer}. First, the target frame and the adjacent reference frame are partitioned into non-overlapping patches by the \emph{Patch Partition layer}. For the consideration of computing speed, the \emph{Patch Partition layer} downsamples the features and restores it to the original resolution at the final stage by the \emph{Pixel Shuffle layer}. Between the \emph{Patch Partition layer} and the \emph{Pixel Shuffle layer}, the spatio-temporal information is aggregated with the \emph{Swin-AutoEncoder}.

\emph{Swin-AutoEncoder} is a Swin Transformer Block (\emph{Swin-TB}) based auto-encoder structure. In \emph{Swin-AutoEncoder}, each patch after segmentation is treated as a token and then calculated the spatio-temporal attention. In the encoder, \emph{Patch Merging Layer} increases the number of channels while features are downsampled, and \emph{Swin-TB} further enhances the features.
For $V_{t}=\left\{X_{t-R}, \ldots, X_{t}, \ldots, X_{t+R}\right\}$, the whole Encoder process can be expressed as
\begin{eqnarray}
E_{1}&=&Estage1\left(V_{t} \right),\\
E_{2}&=&Estage2\left(E_{1} \right),\\
E_{3}&=&Estage3\left(E_{3} \right),
\label{equ:f2}
\end{eqnarray}
where $Estage$ denotes the combination of \emph{Patch Merging} and \emph{Swin-TB}, and 1, 2, 3 represent each stage of the encoder.

Corresponding to the encoder, the decoder uses a \emph{Patch Expanding} Layer to upsample deep features. For each scale, the low-level features of the encoder are connected with the high-level features of the decoder through skip connections to reduce the loss of spatial information caused by downsampling. The whole Encoder process can be expressed as 
\begin{eqnarray}
D_{3}&=&Dstage1\left(E_{3} \right)+E_{3},\\
D_{2}&=&Dstage2\left(D_{3} \right)+E_{2},\\
X_{t}^{m}&=&Dstage3\left(D_{2} \right)+E_{1},
\label{equ:f3}
\end{eqnarray}
where $Dstage$ denotes the combination of \emph{Patch Expanding
} and \emph{Swin-TB}, and 1, 2, 3 represent each stage of the decoder.

By using \emph{Swin-AutoEncoder}, the compressed frames at time $t$ can aggregate the temporal information of the adjacent reference frames and generate a temporal feature map $X_{t}^{m}$.

\subsection{Channel-wise Attention based Quality Enhancement Module (CAQE)}
In order to efficiently fuse the temporal information in each channel and generate residual maps for frames at time $t$, we constructed the CAQE module with efficient channel-level attention. The CAQE module consists of four \emph{Restormer}~\cite{zamir2022restormer} and one \emph{Reconstruct layer}. In which, \emph{Restormer} calculates the channel attention, where temporal information is further fused and enhanced. The final \emph{Reconstruct layer} is a $3\times3$ convolutional layer, which reduces the number of fused feature channels to $1$ to obtain the final residual map. The whole process can be represented as
\begin{eqnarray}
I_{t}&=&Rec\left(Res\left(X_{t}^{m}\right)\right),\\
X_{t}^{e}&=&I_{t}+X_{t},
\label{equ:f4}
\end{eqnarray}
where $Res$ denotes four consecutive stacks of \emph{Restormer} and $Rec$ denotes the final \emph{reconstructed layer}.

As shown in Fig.~\ref{fig:atten}, \emph{Restormer} consists of two parts: \emph{Multi-Dconv Head Transposed Attention} (MDTA) and \emph{Gated-Dconv Feed-Forward Network} (GDFN).
To reduce the computational overhead of the network, MDTA computes the cross-covariance on the channel. Input feature $X_{m}^{t}$, using \emph{Pointwise Convolution} (PW conv) and \emph{Depthwise Convolution} (DW conv) to Generate $\mathbf{Q} \in \mathbb{R}^{H \times W \times C}$, $\mathbf{K} \in \mathbb{R}^{H \times W \times C}$, $\mathbf{V} \in \mathbb{R}^{H \times W \times C}$. Specifically, the PW performs content encoding on the channel and fuses the context information between the channels. DW further encodes spatial context. Then get $\hat{\mathbf{Q}} \in \mathbb{R}^{H W \times C}$, $\hat{\mathbf{K}} \in \mathbb{R}^{H W \times C}$, $\hat{\mathbf{V}} \in \mathbb{R}^{H W \times C}$ through the reshape operation and calculate the dot product of $\hat{\mathbf{Q}}$ and $\hat{\mathbf{K}}$ to generate the channel attention map $\mathbf{M}$ with size $C \times C$. It can be expressed as
\begin{eqnarray}
\mathbf{M}&=&\mathbf{V} \cdot \operatorname{Softmax}(\mathbf{K} \cdot \mathbf{Q})
\label{equ:f5}
\end{eqnarray}
To get more accurate residual information, we utilize GDFN with more complex operations. GDFN adds GELU activation branch and DW on the basis of FN, which can enrich the expression of features and use spatial context information to enhance the recovery of local details.

\subsection{Training Scheme}
For frame $X_{t}$ at time $t$, we use a two-stage training strategy to enhance its quality. In the first stage, we use Charbonnier Loss~\cite{charbonnier1994two} to optimize the parameters of TVQE. In the second stage, we use $\mathcal{L}_{2}$ Loss to further fine-tune the model for a better visual result. Finally, the loss function is defined as

\begin{eqnarray}
\mathcal{L}_{charb}&=&\sqrt{\left({X}^{e}_{t}-{X}^{raw}_{t}\right)^{2} + \epsilon},\\
\mathcal{L}_{mse}&=& \left\|{X}^{e}_{t}-{X}^{raw}_{t}\right\|_{2}^{2},\\
\mathcal{L}&=&\alpha\times \mathcal{L}_{charb}+\beta\times\mathcal{L}_{mse},
\label{equ:loss}
\end{eqnarray}
where ${X}^{e}_{t}$ denotes the enhanced video frame at time $t$, ${X}^{raw}_{t}$ denotes the raw frame, and $\epsilon$ is a constant set to $10^{-6}$. $\alpha$ and $\beta$ are the weights of the loss.

\section{Experiment}
\label{sec:experiment}

\subsection{Datasets}
In this work, we use the MFQE 2.0~\cite{guan2019mfqe} and LDV~\cite{yang2021ntire} datasets for training and JCT-VC~\cite{bossen2013common} dataset for testing.
\subsubsection{MFQE 2.0}
It contains totally $128$ sequences, in which training set contains $108$ sequences. The sequences are acquired from Xiph.org~\cite{montgomery2021xiph} and VQEG~\cite{VGQE}, with resolutions ranging from SIF ($352\times240$) to WQXGA ($2560\times1600$).
\subsubsection{LDV}
It is proposed in NTIRE 2021 challenge~\cite{yang2021ntire1} with $240$ sequences, which consists of training set, validation set and test set. We use $200$ sequences from the training set as training data and $40$ sequences from the validation and test sets for validation, all sequences are $960\times536$ in resolution.
\subsubsection{JCT-VC}
The test set has $18$ sequences, delivered by JCT-VC (Joint Collaborative Team on Video Coding) for evaluating the performance of our model. There are totally five resolutions ranging from $240$p ($416\times240$) to WQXGA ($2560\times1600$), named as Class A to E.

Following~\cite{guan2019mfqe,deng2020spatio}, all sequences were compressed by HEVC HM $16.5$\footnote{https://hevc.hhi.fraunhofer.de/trac/hevc/milestone/HM-16.5} with LDP (Low-Delay-P) configuration. Five QPs (quantization parameters) i.e., $22$, $27$, $32$, $37$,$42$ at different compression bit rates are used for experiments.

\subsection{Implementation Details}
For network structure, the window size is set to $8$ in \emph{Swin-AutoEncoder} from Stage1 to Stage3, the number of \emph{Swin-TB} is [$2$, $2$, $2$], and the attention heads is [$2$, $2$, $2$]. The \emph{Patch embedding} dimension is set to $48$, and the \emph{MLP-ratio} is $1$. The number of \emph{Restormer} in CAQE is set to $4$.

In the training stage, we crop $128\times128$ patches randomly from the compressed video and the corresponding raw video as training samples. Random flips and rotations are also used for data augmentation. The batch size is set to $32$. we train the model using the Adam optimizer ($\beta_{1}$=$0.9$, $\beta_{2}$=$0.999$, $\epsilon$=$10^{-8}$). The learning rate is $10^{-4}$ throughout the training process. In the first stage of training, $\alpha$ is set to $1$ and $\beta$ is set to $0$ in Equ.~(\ref{equ:loss}). In the second stage, $\alpha$ is set to $0$ and $\beta$ is set to $1$. All experiments are performed on the NVIDIA TITAN RTX.

For testing, we evaluate the results using $\Delta$PSNR and $\Delta$SSIM, as well as BD-rate. All tests are performed on the Y-channel in YUV space.

\begin{table*}[htbp]\normalsize
\renewcommand{\arraystretch}{1.2} 
\caption{Quantitative results of $\Delta$PSNR (dB) / $\Delta$SSIM ($\times10^{-2}$) on JCT-VC dataset at 5 different QPs. The best and second best performance are bold and underlined, respectively.}
\label{tab:qrtable}
\begin{threeparttable}
\resizebox{\linewidth}{!}{
\begin{tabular}{c|c|l||ccccccc|c}
\toprule[1.5pt]
QP & \multicolumn{2}{c||}{Approach} & \begin{tabular}[c]{@{}c@{}}AR-CNN\\ \cite{dong2015compression}\end{tabular} & \begin{tabular}[c]{@{}c@{}}DnCNN\\ \cite{zhang2017beyond}\end{tabular} & \begin{tabular}[c]{@{}c@{}}RNAN\\ \cite{zhang2019residual}\end{tabular} & \begin{tabular}[c]{@{}c@{}}MFQE 2.0\\ \cite{guan2019mfqe}\end{tabular} & \begin{tabular}[c]{@{}c@{}}Ding et al.\\ \cite{ding2021patch}\end{tabular} & \begin{tabular}[c]{@{}c@{}}STDF-R3L\\ \cite{deng2020spatio}\end{tabular} & \begin{tabular}[c]{@{}c@{}}RFDA\\ \cite{zhao2021recursive}\end{tabular} &  \begin{tabular}[c]{@{}c@{}}Ours\\ TVQE\end{tabular} \\ \hline
 & \multicolumn{2}{c||}{Metrics} & PSNR / SSIM & PSNR / SSIM & PSNR / SSIM & PSNR / SSIM & PSNR / SSIM & PSNR / SSIM & PSNR / SSIM & PSNR / SSIM \\ \hline
\multirow{19}{*}{37} & \multirow{2}{*}{A} & \textit{Traffic} & 0.24 / 0.47 & 0.24 / 0.57 & 0.40 / 0.86 & 0.59 / 1.02 & \textbf{1.08} / \textbf{1.68} & 0.73 / 1.15 & 0.80 / 1.28 & \underline{0.86} / \underline{1.39} \\
 &  & \textit{PeopleOnStreet} & 0.35 / 0.75 & 0.41 / 0.82 & 0.74 / 1.30 & 0.92 / 1.57 & 0.64 / 1.04 & 1.25 / 1.96 & \underline{1.44} / \underline{2.22} & \textbf{1.46} / \textbf{2.30} \\ \cline{2-11} 
 & \multirow{5}{*}{B} & \textit{Kimono} & 0.22 / 0.65 & 0.24 / 0.75 & 0.33 / 0.98 & 0.55 / 1.18 & 0.69 / 1.36 & 0.85 / 1.61 & \textbf{1.02} / \textbf{1.86} & \underline{0.99} / \underline{1.82} \\
 &  & \textit{ParkScene} & 0.14 / 0.38 & 0.14 / 0.50 & 0.20 / 0.77 & 0.46 / 1.23 & 0.49 / 1.21 & 0.59 / 1.47 & \underline{0.64} / \underline{1.58} & \textbf{0.64} / \textbf{1.63} \\
 &  & \textit{Cactus} & 0.19 / 0.38 & 0.20 / 0.48 & 0.35 / 0.76 & 0.50 / 1.00 & 0.62 / 1.15 & 0.77 / 1.38 & \underline{0.83} / \underline{1.49} & \textbf{0.87} / \textbf{1.59} \\
 &  & \textit{BQTerrace} & 0.20 / 0.28 & 0.20 / 0.38 & 0.42 / 0.84 & 0.40 / 0.67 & 0.50 / 0.87 & 0.63 / 1.06 & \underline{0.65} / \underline{1.06} & \textbf{0.70} / \textbf{1.22} \\
 &  & \textit{BasketballDrive} & 0.23 / 0.55 & 0.25 / 0.58 & 0.43 / 0.92 & 0.47 / 0.83 & 0.60 / 1.04 & 0.75 / 1.23 & \textbf{0.87} / \underline{1.40} & \underline{0.82} / \textbf{1.41} \\ \cline{2-11} 
 & \multirow{4}{*}{C} & \textit{RaceHorses} & 0.22 / 0.43 & 0.25 / 0.65 & 0.39 / 0.99 & 0.39 / 0.80 & 0.40 / 0.88 & 0.55 / 1.35 & \underline{0.48} / \underline{1.23} & \textbf{0.61} / \textbf{1.55} \\
 &  & \textit{BQMall} & 0.28 / 0.68 & 0.28 / 0.68 & 0.45 / 1.15 & 0.62 / 1.20 & 0.74 / 1.44 & 0.99 / 1.80 & \textbf{1.09} / \textbf{1.97} & \underline{1.02} / \underline{1.95} \\
 &  & \textit{PartyScene} & 0.11 / 0.38 & 0.13 / 0.48 & 0.30 / 0.98 & 0.36 / 1.18 & 0.51 / 1.46 & 0.68 / 1.94 & \underline{0.66} / \underline{1.88} & \textbf{0.75} / \textbf{2.14} \\
 &  & \textit{BasketballDrill} & 0.25 / 0.58 & 0.33 / 0.68 & 0.50 / 1.07 & 0.58 / 1.20 & 0.66 / 1.27 & 0.79 / 1.49 & \underline{0.88} / \underline{1.67} & \textbf{0.96} / \textbf{1.91} \\ \cline{2-11} 
 & \multirow{4}{*}{D} & \textit{RaceHorses} & 0.27 / 0.55 & 0.31 / 0.73 & 0.42 / 1.02 & 0.59 / 1.43 & 0.60 / 1.44 & 0.83 / 2.08 & \underline{0.85} / \underline{2.21} & \textbf{0.87} / \textbf{2.27} \\
 &  & \textit{BQSquare} & 0.08 / 0.08 & 0.13 / 0.18 & 0.32 / 0.63 & 0.34 / 0.65 & 0.79 / 1.14 & 0.94 / 1.25 & \underline{1.05} / \underline{1.39} & \textbf{1.16} / \textbf{1.62} \\
 &  & \textit{BlowingBubbles} & 0.16 / 0.35 & 0.18 / 0.58 & 0.31 / 1.08 & 0.53 / 1.70 & 0.62 / 1.95 & 0.74 / 2.26 & \underline{0.78} / \underline{2.40} & \textbf{0.81} / \textbf{2.48} \\
 &  & \textit{BasketballPass} & 0.26 / 0.58 & 0.31 / 0.75 & 0.46 / 1.08 & 0.73 / 1.55 & 0.85 /1.75 & 1.08 / 2.12 & \textbf{1.12} / \underline{2.23} & \underline{1.07} / \textbf{2.31} \\ \cline{2-11} 
 & \multirow{3}{*}{E} & \textit{FourPeople} & 0.37 / 0.50 & 0.39 / 0.60 & 0.70 / 0.97 & 0.73 / 0.95 & 0.95 / 1.12 & 0.94 / 1.17 & \underline{1.13} / \underline{1.36} & \textbf{1.20} / \textbf{1.44} \\
 &  & \textit{Johnny} & 0.25 / 0.10 & 0.32 / 0.40 & 0.56 / 0.88 & 0.60 / 0.68 & 0.75 / 0.85 & 0.81 / 0.88 & \underline{0.90} / \underline{0.94} & \textbf{1.10} / \textbf{1.31} \\
 &  & \textit{KristenAndSara} & 0.41 / 0.50 & 0.42 / 0.60 & 0.63 / 0.80 & 0.75 / 0.85 & 0.93 / 0.91 & 0.97 / 0.96 & \underline{1.19} / \underline{1.15} &\textbf{1.27} / \textbf{1.21} \\ \cline{2-11} 
 & \multicolumn{2}{c||}{Average} & 0.23 / 0.45 & 0.26 / 0.58 & 0.44 / 0.95 & 0.56 / 1.09 & 0.69 / 1.25 & 0.83 / 1.51 & \underline{0.91} / \underline{1.62} & \textbf{0.95} / \textbf{1.75} \\\bottomrule[1.5pt]
 \toprule[1.5pt]
42 & \multicolumn{2}{c||}{Average} & 0.29 / 0.96 & 0.22 / 0.77 & – / – & 0.59 / 1.65 & 0.69 / 1.86 & – / – & \underline{0.82} / \underline{2.20} & \textbf{0.97} / \textbf{2.58} \\ \hline
32 & \multicolumn{2}{c||}{Average} & 0.18 / 0.19 & 0.26 / 0.35 & 0.41 / 0.62 & 0.52 / 0.68 & 0.67 / 0.83 & 0.86 / 1.04 & \underline{0.87} / \underline{1.07} & \textbf{0.91} / \textbf{1.17} \\\hline
27 & \multicolumn{2}{c||}{Average} & 0.18 / 0.14 & 0.27 / 0.24 & – / – & 0.49 / 0.42 & 0.63 / 0.52 & 0.72 / 0.57 & \underline{0.82} / \underline{0.68} & \textbf{0.85} / \textbf{0.74}  \\\hline
22 & \multicolumn{2}{c||}{Average} & 0.14 / 0.08 & 0.29 / 0.18 & – / – & 0.46 / 0.27 & 0.55 / 0.29 & 0.63 / 0.34 & \textbf{0.76} / \underline{0.42} & \underline{0.74} / \textbf{0.42} \\\bottomrule[1.5pt]
\end{tabular}}
\begin{tablenotes}
\footnotesize
\item Video resolution: Class A ($2560 \times 1600$), Class B ($1920 \times 1080$), Class C ($832 \times 480$), Class D ($480 \times 240$), Class E ($1280 \times 720$)
\end{tablenotes}
\end{threeparttable}
\end{table*}


\begin{table}[htbp]
\renewcommand{\arraystretch}{1.2} 
\caption{Comparison of inferred speed and GPU consumption between our method and some mainstream methods. For a fair comparison, all methods were retested on the NVIDIA TITAN RTX. Results are reported in frames per second (FPS) and Test Memory (GB). The best and second best performance are bold and underlined, respectively.}
\label{tab:my-table}
\resizebox{\linewidth}{!}{
\begin{tabular}{l||cccc}
\toprule[1pt]
\multicolumn{1}{l||}{\multirow{2}{*}{Method}} & \multicolumn{4}{c}{Test Memory(GB) / FPS} \\ \cline{2-5} 
\multicolumn{1}{l||}{} & 240p & 480p & 720p & 1080p \\\hline
STDF-R3L & 1.2 / \textbf{42.5} & 1.9 / \textbf{13.7} & 3.1 / \textbf{6.4} & \;\;5.8 / \textbf{3.1} \\\hline
RFDA & 1.5 / 32.1 & 2.9 / 11.2 & 5.4 / 5.1 & 10.5 / 2.1 \\\hline
TVQE & \textbf{1.1} / \underline{35.9} & \textbf{1.6} / \underline{12.2} & \textbf{2.5} / \underline{5.5} & \;\;\textbf{4.7} / \underline{2.6}\\
\bottomrule[1pt]
\end{tabular}}
\end{table}

\subsection{Comparison with State of the Art Methods}
To demonstrate the effectiveness of our method, we compare the proposed method with seven state-of-the-art methods, including single-frame based methods (AR-CNN~\cite{dong2015compression}, DnCNN~\cite{zhang2017beyond}, RNAN~\cite{zhang2019residual}) and multi-frame based methods (MFQE $2.0$~\cite{guan2019mfqe}, ~\cite{ding2021patch}, STDF-R$3$L~\cite{deng2020spatio} and RFDA~\cite{zhao2021recursive}.)

\subsubsection{Quantitative Results}
Table~\ref{tab:qrtable} presents the quantitative results of our method and seven state-of-the-art methods on $\Delta$PSNR and $\Delta$SSIM. It can be seen that our method TVQE outperforms the seven models in terms of the average $\Delta$PSNR at four QPs and in terms of the average $\Delta$SSIM at all QPs. Meanwhile, the gain of our method TVQE on SSIM is higher than PSNR obviously. Such as QP=$42$, our method TVQE gains $9.7\%$ on $\Delta$PSNR and $11.8\%$ on $\Delta$SSIM over the second best method RFDA, which indicates that our method TVQE provides better visual effects. More specifically, our method outperforms the current state-of-the-art methods on most sequences when QP=$37$. 

As for the single-frame based methods, RNAN proposes non-local attention blocks to obtain the remote dependence of the feature map, and finally achieves the best performance among all single-frame based methods. RNAN gains about $69\%$ over DnCNN, which reflects the superiority of the transformer-based method. 
However, the single-frame based method cannot use temporal information and has limited performance. Our method computes spatial attention and channel attention over multiple frames, and achieves $\Delta$PSNR of $0.95$, which is about $116\%$ compared to RNAN. 

As for the multi-frame based methods, MFQE $2.0$ calculates the explicit optical flow of compressed video and achieves an average $\Delta$PSNR of $0.56$. STDF proposes deformable convolution to align video frames, which solves the problem of inaccurate optical flow estimation of MFQE $2.0$, and achieves an average $\Delta$PSNR of $0.83$. RFDA utilizes the RF (Recursive Fusion) module to exploit temporal information within a longer time range, and obtains $\Delta$PSNR of $0.91$. Our method TVQE utilizes the long-range modeling property of Transformer to exploit the temporal information, which further increases the PSNR with an average $\Delta$PSNR of $0.96$, which demonstrates the effectiveness of our method.

\begin{table*}[htbp]
\renewcommand{\arraystretch}{1.2} 
\centering
\caption{Results of BD-Bate reduction (\%) at QP = $22, 27, 32, 37$ and $42$ with the HEVC as anchor. The best and second best performance are in bold and underlined, respectively.}
\label{tab:bdrate}
\begin{tabular}{ll|cccccccccc}
\toprule[1.5pt]
\multicolumn{2}{c||}{Sequence} & \begin{tabular}[c]{@{}c@{}}AR-CNN\\\cite{dong2015compression}\end{tabular} & \begin{tabular}[c]{@{}c@{}}DnCNN\\ \cite{zhang2017beyond}\end{tabular} & \begin{tabular}[c]{@{}c@{}}Li et al.\\\cite{li2017efficient}\end{tabular} & \begin{tabular}[c]{@{}c@{}}DCAD\\\cite{wang2017novel}\end{tabular} & \begin{tabular}[c]{@{}c@{}}DS-CNN\\\cite{yang2018enhancing}\end{tabular} & \begin{tabular}[c]{@{}c@{}}MFQE 1.0\\\cite{yang2018multi}\end{tabular} & \begin{tabular}[c]{@{}c@{}}MFQE 2.0\\\cite{guan2019mfqe}\end{tabular} & \begin{tabular}[c]{@{}c@{}}STDF-R3L\\ \cite{deng2020spatio}\end{tabular} & \begin{tabular}[c]{@{}c@{}}RFDA\\\cite{zhao2021recursive}\end{tabular} & \begin{tabular}[c]{@{}c@{}}Ours\\ TVQE\end{tabular} \\ \hline
\multicolumn{1}{l|}{\multirow{2}{*}{A}} & \multicolumn{1}{l||}{\textit{Traffic}} & 7.40 & 8.54 & 10.08 & 9.97 & 9.18 & 14.56 & 16.98 & 21.19 & \underline{22.70} & \textbf{24.00} \\
\multicolumn{1}{l|}{} & \multicolumn{1}{l||}{\textit{PeopleOnStreet}} & 6.99 & 8.28 & 9.64 & 9.68 & 8.67 & 13.71 & 15.08 & 17.42 & \underline{21.11} & \textbf{22.86}  \\ \hline
\multicolumn{1}{l|}{\multirow{5}{*}{B}} & \multicolumn{1}{l||}{\textit{Kimono}} & 6.07 & 7.33 & 8.51 & 8.44 & 7.81 & 12.60 & 13.34 & 17.96 & \underline{22.32} & \textbf{22.58}  \\
\multicolumn{1}{l|}{} & \multicolumn{1}{l||}{\textit{ParkScene}} & 4.47 & 5.04 & 5.35 & 5.68 & 5.42 & 12.04 & 13.66 & 18.1 & \textbf{19.85} & \underline{19.78}  \\
\multicolumn{1}{l|}{} & \multicolumn{1}{l||}{\textit{Cactus}} & 6.16 & 6.80 & 8.23 & 8.69 & 8.78 & 12.78 & 14.84 & 21.54 & \underline{21.78} & \textbf{23.77}  \\
\multicolumn{1}{l|}{} & \multicolumn{1}{l||}{\textit{BQTerrace}} & 6.86 & 7.62 & 8.79 & 9.98 & 8.67 & 10.95 & 14.72 & 24.71 & \underline{24.41} & \textbf{26.89}  \\
\multicolumn{1}{l|}{} & \multicolumn{1}{l||}{\textit{BasketballDrive}} & 5.83 & 7.33 & 8.61 & 8.94 & 7.89 & 10.54 & 11.85 & 16.75 & \underline{20.24} & \textbf{20.70}  \\ \hline
\multicolumn{1}{l|}{\multirow{4}{*}{C}} & \multicolumn{1}{l||}{\textit{RaceHorses}} & 5.07 & 6.77 & 7.10 & 7.62 & 7.48 & 8.83 & 9.61 & 15.62 & \textbf{14.29} & \underline{13.67}  \\
\multicolumn{1}{l|}{} & \multicolumn{1}{l||}{\textit{BQMall}} & 5.60 & 7.01 & 7.79 & 8.65 & 7.64 & 11.11 & 13.50 & 21.12 & \textbf{21.62} & \underline{20.81}  \\
\multicolumn{1}{l|}{} & \multicolumn{1}{l||}{\textit{PartyScene}} & 1.88 & 4.02 & 3.78 & 4.88 & 4.08 & 6.67 & 11.28 & 22.24 & \underline{21.11} & \textbf{22.28} \\
\multicolumn{1}{l|}{} & \multicolumn{1}{l||}{\textit{BasketballDrill}} & 4.67 & 8.02 & 8.66 & 9.8 & 8.22 & 10.47 & 12.63 & 15.94 & \underline{18.06} & \textbf{20.82}  \\ \hline
\multicolumn{1}{l|}{\multirow{4}{*}{D}} & \multicolumn{1}{l||}{\textit{RaceHorses}} & 5.61 & 7.22 & 7.68 & 8.16 & 7.35 & 10.41 & 11.55 & 15.26 & \textbf{17.57} & \underline{17.03}  \\
\multicolumn{1}{l|}{} & \multicolumn{1}{l||}{\textit{BQSquare}} & 0.68 & 4.59 & 3.59 & 6.11 & 3.94 & 2.72 & 11.00 & 33.36 & \underline{31.65} & \textbf{33.50}  \\
\multicolumn{1}{l|}{} & \multicolumn{1}{l||}{\textit{BlowingBubbles}} & 3.19 & 5.10 & 5.41 & 6.13 & 5.55 & 10.73 & 15.2 & 23.54 & \underline{22.89} & \textbf{23.62}  \\
\multicolumn{1}{l|}{} & \multicolumn{1}{l||}{\textit{BasketballPass}} & 5.11 & 7.03 & 7.78 & 8.35 & 7.49 & 11.70 & 13.43 & 18.42 & \textbf{20.42} & \underline{20.12}  \\ \hline
\multicolumn{1}{l|}{\multirow{3}{*}{E}} & \multicolumn{1}{l||}{\textit{FourPeople}} & 8.42 & 10.12 & 11.46 & 12.21 & 11.13 & 14.89 & 17.50 & 22.91 & \underline{22.84} & \textbf{25.33}  \\
\multicolumn{1}{l|}{} & \multicolumn{1}{l||}{\textit{Johnny}} & 7.66 & 10.91 & 13.05 & 13.71 & 12.19 & 15.94 & 18.57 & 24.55 & \underline{23.87} & \textbf{28.99}  \\
\multicolumn{1}{l|}{} & \multicolumn{1}{l||}{\textit{KristenAndSara}} & 8.94 & 10.65 & 12.04 & 12.93 & 11.49 & 15.06 & 18.34 & 23.64 & \underline{24.47} & \textbf{27.95}  \\ \hline
\multicolumn{2}{c||}{Average} & 5.59 & 7.36 & 8.20 & 8.89 & 7.85 & 11.41 & 14.06 & 20.79 & \underline{21.73} & \textbf{23.04} \\\bottomrule[1.5pt]
\end{tabular}
\end{table*}

\subsubsection{Speed and Cost Comparison}
Table~\ref{tab:my-table} shows the inference speed and GPU consumption of our method, compared to STDF-R$3$L~\cite{ding2021patch} and RFDA~\cite{deng2020spatio}. As can be seen, although our model is based on Transformer, it still has a fast inference speed. At the same time, our method is hardware friendly as it requires less GPU memory. More specifically, comparing to STDF at $1080$p resolution, our method is $16.1\%$ slower at inference speed (from $3.1$ to $2.6$, see Table~\ref{tab:my-table}), but with a $19.0\%$ reduction in memory consumption (from $5.8$ to $4.7$, see Table~\ref{tab:my-table}), as well as with a $14.5\%$ improvement in average $\Delta$PSNR performance at QP$37$ (from $0.83$ to $0.95$, see Table~\ref{tab:qrtable}). RFDA is based on STDF by adding RF module, and thus consumes more GPU resources. Comparing to RFDA, our method outperforms RFDA in terms of inference speed and GPU consumption at all resolutions. Specifically, under $1080$p resolution, the inference speed is improved by $23.8\%$ (from $2.1$ to $2.6$, see Table~\ref{tab:my-table}) and GPU consumption is reduced by $55.2\%$ (from $10.5$ to $4.7$, see Table~\ref{tab:my-table}).

\subsubsection{Quality Fluctuation}

 \begin{figure}[htbp]
  \centerline{\includegraphics[width=9cm]{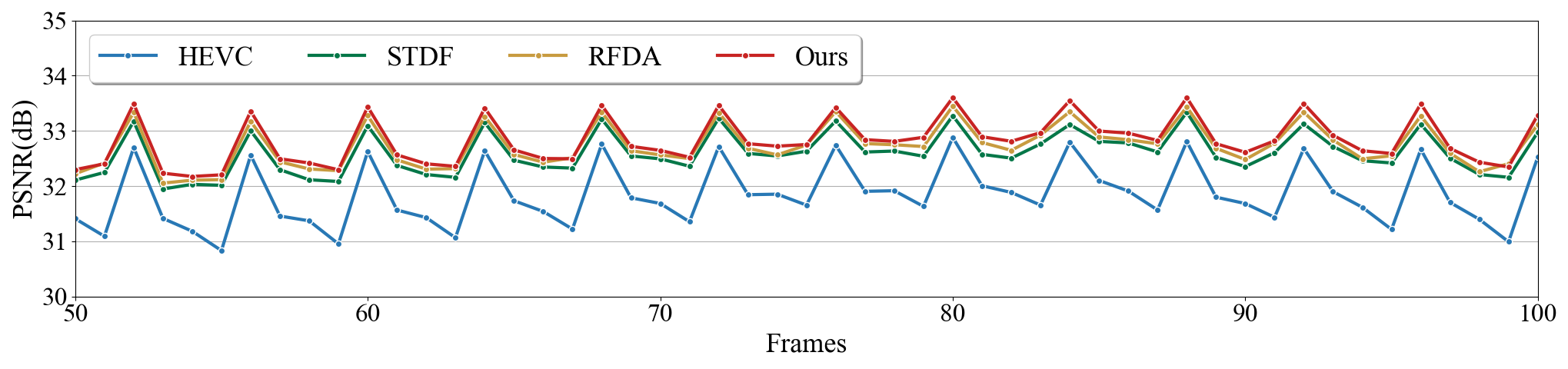}}
  \centerline{\includegraphics[width=9cm]{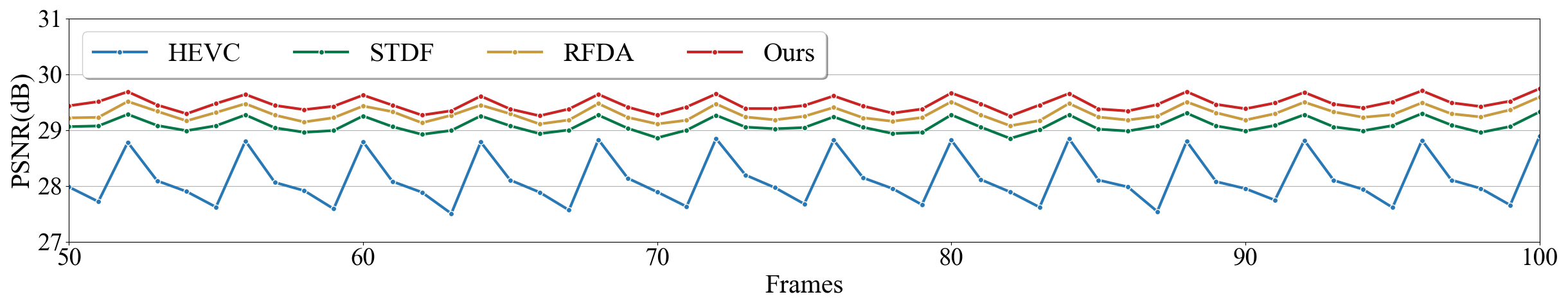}}
\caption{Illustration of quality fluctuations of two sequences. (Top: Class C, \emph{BasketballDrill}. Bottom: Class D,  \emph{BQSquare}.)}
\label{fig:zlbd}
\end{figure}

The PSNR of each frame in two sequences are plotted in Fig.~\ref{fig:zlbd}. It can be observed that the HEVC compressed sequences have severe quality fluctuations (i.e., quality differences between high quality frames and adjacent low quality frames). Compared to both STDF and RFDA, our method provides better PSNR and smaller quality fluctuations, effectively improving the QoE.

\subsubsection{Rate-Distortion performance}
\begin{figure}[htbp]
    \centering\includegraphics[width=4.3cm]{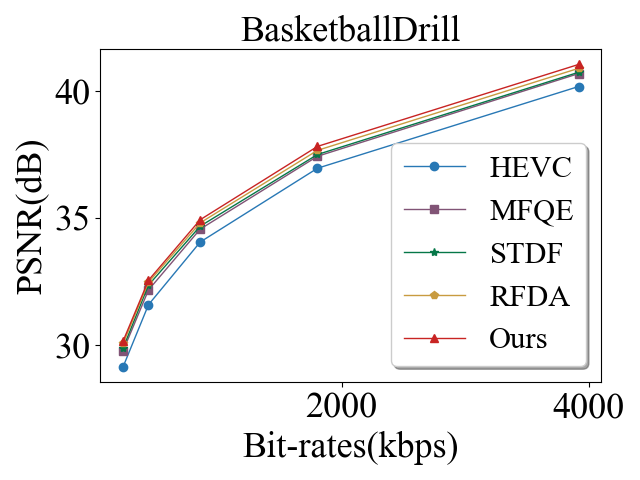}
    \centering\includegraphics[width=4.4cm]{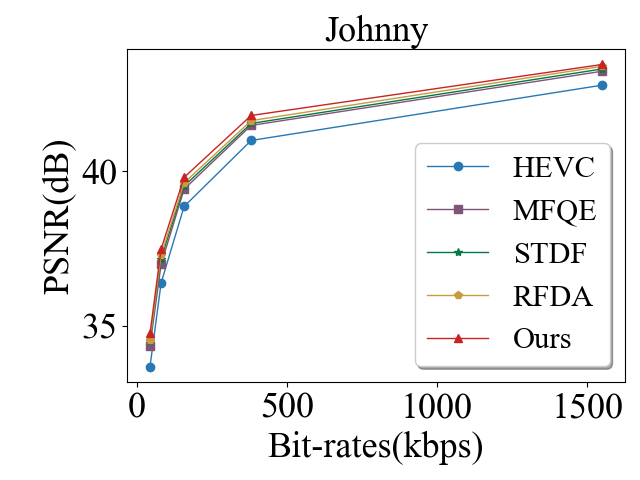}\\
    \centering\includegraphics[width=4.3cm]{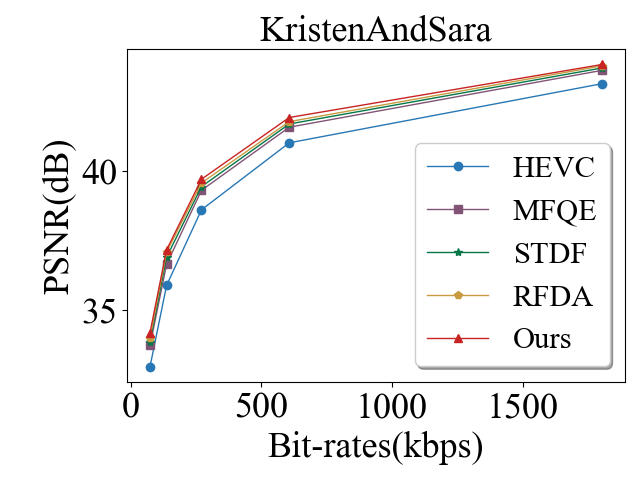}
    \centering\includegraphics[width=4.3cm]{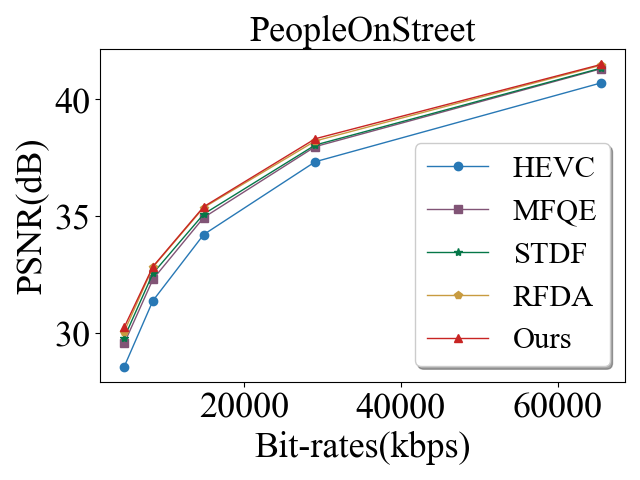}
\caption{Rate-Distortion curves of sequences BasketballDrill, Johnny, KristenAndSara and PeopleOnStreet.}
\label{fig:bdrate}
\end{figure}

Fig.~\ref{fig:bdrate} presents the rate distortion curves for the four sequences. It can be seen that our method outperforms other methods on both sequences with huge motion (e.g., Class C, \emph{BasketballDrill}) and smooth motion (e.g., Class E, \emph{Johnny}). In addition, we also calculate the BD-rate reduction of PSNR on five QPs (= $22$, $27$, $32$, $37$, $42$). As shown in Table~\ref{tab:bdrate}, our method provides an average BD-rate reduction of $23.04\%$, which is better than the state-of-the-art CNN method RFDA with $21.73\%$. It demonstrates that our method exhibits a better rate distortion performance, which can provide superior visual effects with the same compression rate.

 \begin{figure*}[htbp]
 \centerline{\includegraphics[width=18cm]{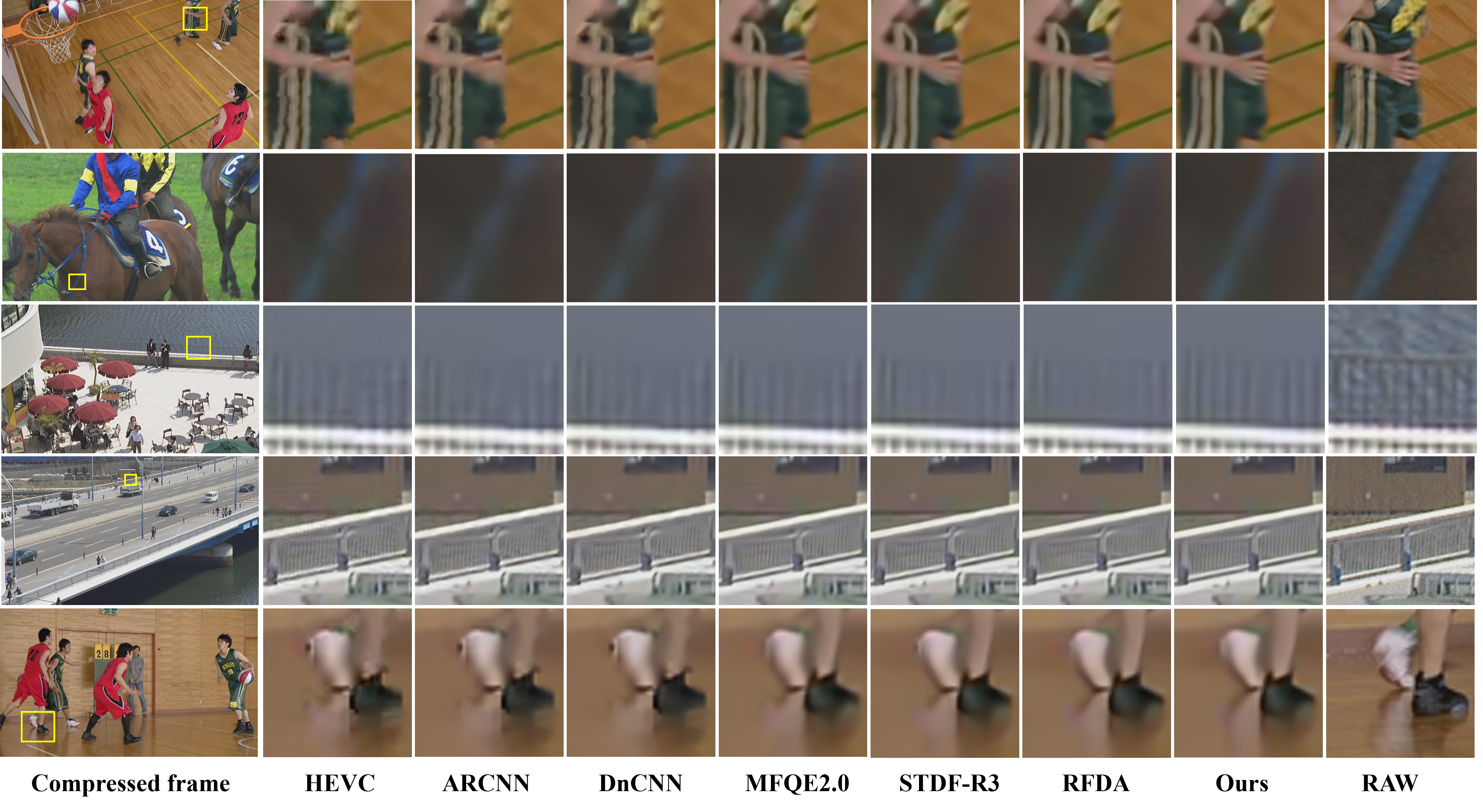}}
\caption{Qualitative results at QP $37$. Video from top to bottom: \emph{BasketballDrill}, \emph{Racehorses}, \emph{BQSquare}, \emph{BQTerrace}, \emph{BasketballPass}. For a fair comparison, for each method we only enhance on Y component.}
\label{fig:Qualitative}
\end{figure*}

\subsubsection{Qualitative Results}
Fig.~\ref{fig:Qualitative} gives the qualitative results for the five sequences. It can be seen that the single-frame based quality enhancement methods~\cite{dong2015compression,zhang2017beyond} do not make use of temporal information, so the enhanced video frames still have serious compression artifacts (e.g., block effect, ringing effect). With the help of temporal information, CNN-based multi-frame quality enhancement methods~\cite{guan2019mfqe,deng2020spatio} provide better visual effects with the help of reference frames, but the locality of the convolution operation prevents these methods from taking full advantage of the temporal information of reference frames, resulting in enhanced video frames that are too smooth and lack of detailed texture. RFDA~\cite{zhao2021recursive} further implicitly expands the temporal range with RF to better recover details, but the RF module consumes large computational resources and decreases the inference speed. Our proposed method TVQE is based on transformer, which has better remote correlation capability than convolution, thus resulting in better exploration of spatio-temporal information and finer recovery of textures. For example in Fig.~\ref{fig:Qualitative}, the player's fingers in \emph{BaskerballDrill}, the rope on the horse in \emph{Racehorses}, the textures on the railings in \emph{BQSquare} and \emph{BQTerrace}, and the shadow of the sneaker in \emph{BaskerballPass} can be better recovered by our method than other methods.

\subsection{Ablation Study}
In this section, we perform ablation experiments as well as specific analysis of the proposed method. We take STDF-R3L as baseline and replace the modules from different models to analyze their effects. For a fair comparison, all models are retrained by the same training approach as the proposed method. The results of inference speed FPS and GPU consumption are obtained at 1080p resolution, and $\Delta$PSNR / $\Delta$SSIM takes the average results of the test sequences in ClassA-E at QP=$37$.

\begin{table}[htbp]
\renewcommand{\arraystretch}{1.2} 
\caption{Ablation study on SSTF and CAQE. The best and second best performance are bold and underlined, respectively.}
\label{tab:withqe}
\resizebox{\linewidth}{!}{
\begin{tabular}{l||ccc}
\toprule[1pt]
Method & $\Delta$PSNR / $\Delta$SSIM & Test Memory(GB) & FPS \\
\hline
STDF+QE & 0.83 / 1.51 & \underline{5.8} & \underline{3.1} \\
STFF+QE & \underline{0.91} / \underline{1.62} & 10.1 & 2.4 \\
SSTF+QE & \textbf{0.91} / \textbf{1.71} & \textbf{5.3} & \textbf{3.4}\\\hline
STDF+CAQE & 0.86 / 1.56 & \underline{5.1} & \underline{2.3} \\
STFF+CAQE & \underline{0.93} / \underline{1.65} & 9.4 & 1.6\\
SSTF+CAQE & \textbf{0.95} / \textbf{1.75} & \textbf{4.7} & \textbf{2.6}\\
\bottomrule[1pt]
\end{tabular}}
\end{table}

\subsubsection{Effectiveness of SSTF}
To illustrate the effectiveness of the SSTF module, we compare the proposed SSTF with the baseline STDF and RFDA. As shown in Table~\ref{tab:withqe}, by replacing the STDF module and the STFF in RFDA with the SSTF (from the first to the third row), the SSTF provides a larger performance improvement while having a faster inference speed and lower GPU consumption. Specifically, compared to the baseline STDF, the Transformer-based SSTF is able to explore global temporal information within a time window, with an improvement of $\Delta$PSNR by $0.08$ (from $0.83$ to $0.91$) and $\Delta$SSIM ($\times10^{-2}$) by $0.2$ (from $1.51$ to $1.71$). Moreover, benefit from the \emph{Swin-AutoEncoder} structure and skip connections, SSTF has a $9.7\%$ speedup (from $3.1$ to $3.4$) inference speed compared to STDF and $8.6\%$ reduction in GPU consumption (from $5.8$ to $5.3$). Compared with STFF, SSTF does not have to utilize additional information outside the time window, resulting in $47.5\%$ lower GPU consumption (from $10.1$ to $5.3$) and $41.6\%$ higher inference express (from $2.4$ to $3.4$), which demonstrates the effectiveness of the SSTF module.

\subsubsection{Effectiveness of CAQE}
To illustrate the effectiveness of the CAQE module, we replace the quality enhancement module in the baseline STDF, RFDA and this method with CAQE (fourth to sixth rows). CAQE calculates the channel attention and effectively fuses the temporal information between channels. In terms of the results of this method (third and sixth lines), CAQE provides a $\Delta$PSNR gain of $0.04$ (from $0.91$ to $0.95$) and a $\Delta$SSIM gain of $0.04$ (from $1.71$ to $1.75$) compared to the QE with CNN structure. Meanwhile, the channel-level attention resulted in a further $11.3\%$ reduction in memory consumption (from $5.3$ to $4.7$) and the inference speed was reduced from $3.4$ to $2.6$, but the overall inference speed was still better than RFDA ($2.1$, see Table~\ref{tab:my-table}), reflecting the effectiveness of the CAQE module.
\section{Conclusion}
\label{sec:con}
In this paper, we propose an end-to-end transformer based network TVQE for compressed video enhancement, which mainly consists of two modules, SSTF module and CAQE module. SSTF module can efficiently explore temporal information within the time window, while CAQE can well fuses the temporal information. The proposed method outperforms the CNN-based methods in terms of performance, inference speed and GPU consumption. The proposed module can also be used in other fields, such as video super-resolution and video interpolation, to explore and fuse temporal information more effectively.


%





\ifCLASSOPTIONcaptionsoff
  \newpage
\fi


\bibliographystyle{IEEEbib}
\bibliography{ref}
\end{document}